\documentclass[journal=jacsat,manuscript=article]{achemso}

\usepackage[version=3]{mhchem} 
\usepackage{amsmath}
\usepackage{amssymb}

\author{Benedikt O. Birgisson}
\affiliation{Science Institute and Faculty of Physical Sciences, U. of Iceland, 107 Reykjavík, Iceland}
\author{Marta Gałyńska}
\affiliation{Science Institute and Faculty of Physical Sciences, U. of Iceland, 107 Reykjavík, Iceland}
\alsoaffiliation{Faculty of Chemistry, Nicolaus Copernicus University in Toru\'n, 87-100 Toru\'n, Poland}
\author{Hemanadhan Myneni}
\affiliation{Science Institute and Faculty of Physical Sciences, U. of Iceland, 107 Reykjavík, Iceland}
\author{Elvar Ö. Jónsson}
\affiliation{Science Institute and Faculty of Physical Sciences, U. of Iceland, 107 Reykjavík, Iceland}
\author{Ragnar Bjornsson}
\affiliation{Laboratoire Chimie et Biologie des Métaux, F-38054 Grenoble, France}
\author{Hannes Jónsson}
\email{hj@hi.is }
\affiliation{Science Institute and Faculty of Physical Sciences, U. of Iceland, 107 Reykjavík, Iceland}
\title{
Localised and Delocalised Charge Distribution in a Diamine Cation and Rydberg Excited State:
A Challenging Test for Density Functionals
}

\begin{document}






\begin{tocentry}
\includegraphics[width =0.95 \textwidth]{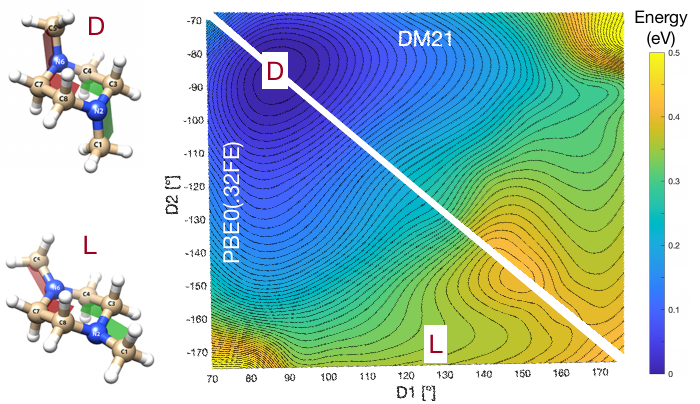}
\end{tocentry}

\begin{abstract}
The balance between localised and delocalised electron distribution in the N,N’-dimethylpiperazine (DMP) molecule in the 3s Rydberg excited state and in the fully ionised DMP$^+$ provides a valuable test of density functionals, in particular the weight of Fock exchange (FE) in hybrid functionals and the scaling of explicit orbital-based self-interaction correction (SIC) applied to less elaborate functionals.  
We present results of calculations using density functionals of all rungs of Jacob's ladder, ranging from LDA to the DM21 machine learned local hybrid as well as double hybrids. 
For DMP$^+$, the commonly used hybrid functionals, such as PBE0(.25) with FE weight of 0.25, as well as PBE0(.32) with a weight of 0.32 only produce the more delocalised charge. The latter mimics the DM21 energy surface well. However, hybrid functionals with stronger FE, such as 
BHLYP
and PBE0(.50) as well as some double hybrid functionals, also produce local energy minima corresponding to localised charge. When full SIC is applied to the PBE functional, an energy surface analogous to hybrid functionals with FE weight of 0.50 is obtained, while the scaling of SIC by 0.50, which has previously been shown to give improved atomisation energy 
and band gap of solids, 
only produces the more delocalised charge.
For the 3s Rydberg excited state of DMP, two types of configurations with a localised hole are obtained in calculations using the PBE0(.32) functional, in addition to the delocalised hole, but only the latter is found with the PBE functional, showing that a significant reduction of the self-interaction error is needed in order to obtain agreement between density functional calculations and experimental measurements of the Rydberg state.
\end{abstract}




The need for accurate description of the relative energy of a localised and a delocalised electron distribution is a challenge that often arises in the theoretical description of molecules, solid materials and biological systems. In mixed-valence compounds, for example, where two equivalent or near-equivalent redox sites are present, ionisation can result in the formation of either a localised or a delocalised electronic distribution.\cite{demadis2001,hankache2011,heckmann2012,blondin1990,solomon} 
Accurate description of these phenomena in electronic structure calculations has turned out to be a significant challenge, revealing deficiencies in both wave function theory (WFT) and density functional theory (DFT) based approaches. In commonly used implementations of DFT, i.e. Kohn-Sham functionals, an overemphasis on delocalisation is often present.\cite{cohen2012,johnson2022}
The semi-local functional form for the exchange-and-correlation term leads to incomplete cancellation of the non-local self-interaction error in the Kohn-Sham estimate of the classical Coulomb interaction, the Hartree term, since it is based only on the total electron density rather than orbital densities. 
In Hartree-Fock (HF) calculations, however, where the self-interaction in the Hartree term gets cancelled out by the self-interaction in the infinite range, non-local Fock exchange (FE), there is an overemphasis on localised states. 
Hybrid functionals, where FE evaluated using the Kohn-Sham orbitals is included with some weight coefficient, can improve the balance between localised and delocalised states but the calculated results depend strongly on the value chosen for this parameter. 
The most commonly used hybrid functionals, with a weight of 0.20 to 0.25, can still overemphasise delocalisation.  A common pragmatic way of dealing with this problem is to increase the weight of FE. For some systems a value of around 0.35 or even larger has been found to be necessary in order to describe correctly molecules with localised electron distribution.\cite{kaupp_review2014} 
In the original hybrid functional
\cite{BHLYP} where the weight coefficient was derived approximately using the adiabatic connection formula, the FE weight is even larger, 0.50.  
However, too large value can lead to an erroneously localised description of intrinsically delocalised electron distribution. 

DFT functionals can be tested against high-level WFT results for small enough molecules.
Kaupp and coworkers have discussed several inorganic\cite{kaup_inorg0,kaup_inorg1} and organic mixed-valence systems\cite{kaup_org0,kaup_org1,kaup_org2,kaup_org3,kaup_org4} that are challenging for DFT approaches. 
 Overall, global hybrid functionals with a factor of 0.35 to 0.45 for FE, as well as range-separated and local hybrid functionals, have been found to give good agreement with high-level WFT calculations in many cases. 
Range-separated hybrid DFT functionals have recently been shown to describe delocalisation well in 
some
organic mixed valence molecules.\cite{bredas2014} 

A less explored way of dealing with the problem is to apply explicit self-interaction correction such as the one proposed by Perdew and Zunger (PZ-SIC)\cite{sic}. There, a correction is made for each orbital separately and thus it represents an independent electron approach and the full correction is found to lead to results similar to HF calculations in many cases, such as for the atomisation energy of molecules.\cite{Klupfel12}
A scaling of the PZ-SIC by a half based on a justification analogous to the adiabatic connection argument used in the derivation of the 
original
hybrid functional\cite{BHLYP} has been shown to give good estimates of atomisation energy as well as band gaps of solids.\cite{Jonsson11,Klupfel12}
However, in order to obtain the correct -$1/r$ long range form of the effective potential, which is, in particular, important for the description of the diffuse electron in a Rydberg state of a molecule or a loosely bound anion, the full PZ-SIC is required.\cite{Gudmundsdottir13,Zhang16} 
PZ-SIC has been implemented in a variational, self-consistent way with complex valued orbitals\cite{Lehtola16,Ivanov21a,Ivanov21b,gpaw2} 
and shown to give good results in studies of several molecules and solids, in particular where the delocalisation error is significant in commonly used Kohn-Sham functionals.\cite{Gudmundsdottir15}

In a recent effort, machine learning has been used to construct density functionals of the local hybrid form where, among other things, data on the balance between localised and delocalised states is included in the training.\cite{dm21}
This deep mind functional, referred to as DM21, is generated by training a neural network with extensive molecular data, in particular where fractional charge and spin problems occur in regular Kohn-Sham functionals. DM21 has been shown to describe well main group thermochemistry and systems exhibiting energy or density errors due to self-interaction error.
A few variants have also been developed with different training sets, namely DM21m, DM21mc and DM21mu.


The radical cation of N,N’-dimethylpiperazine (DMP) dissolved in acetonitrile has been studied by various spectroscopic techniques and the spin density found to be  
distributed over both N atoms as well as the C-C bonds in a through-$\sigma$-bond interaction.\cite{Brouwer1994,Brouwer1998}
Time-resolved pump/probe spectra of gas phase DMP where the molecule is first excited to a 3p Rydberg state and then quickly relaxes to a 3s Rydberg state show that a localised hole is initially formed, followed by a conversion to a more stable state where the hole is distributed and the binding energy for the Rydberg electron 
(cation energy minus Rydberg state energy)
 is lower.\cite{Sanghamitra13,dmp-nat}
In a Rydberg state, an electron is excited into a diffuse orbital with large spatial extent and provides, to first approximation, a nearly uniform background charge for the molecular ion core.
The question then arises whether a localised state similarly exists for the gas phase cation where an electron has been fully removed.
The molecular structure corresponding to localised and delocalised charge of the cation has been calculated\cite{dmp0} using the BHLYP 
functional and is shown in figure 1. 
The atomic coordinates differ significantly between the two and they, thereby, correspond to different minima on the energy surface.
The Rydberg state energy surface is often assumed to be nearly identical to that of the corresponding cation, consistent with the narrowness of peaks in the ionisation spectra of Rydberg excited molecules.\cite{Stankus19,Yong21} 
The peaks are narrow because of little vibrational broadening, indicating that the energy surfaces of the Rydberg excited molecule and cation have similar shape. 
The existence of a localised hole for the Rydberg excited molecule has, therefore, been taken to indicate the existence of a localised charge distribution in the cation.\cite{dmp-nat}
More recently, ultrafast X-ray scattering has been used to study DMP in the 3s Rydberg excited state and the 
data analysed to determine the atomic structure as a function of time after excitation.\cite{Yong21} 
The structure at early time was found to be close to the one calculated for the localised charge distribution of the cation.\cite{dmp0}

Calculations using several electronic structure methods do not, however, produce an energy minimum corresponding to localised charge on the DMP$^+$ energy surface, suggesting the electronic structure methods used are not accurate enough or, alternatively, that the Rydberg state energy surface is significantly different from that of the cation.  
The BHLYP functional
predicts the presence of energy minima corresponding to localised charge, while most commonly used density functionals do not.\cite{dmp-nat}
When full PZ-SIC is applied to the PBE functional,\cite{pbe_0,pbe_1} as is needed to accurately describe  
the highly diffuse Rydberg orbital,
such energy minima are also produced.
Some trusted WFT methods, such as the single reference coupled cluster CCSD(T) 
and CCSDT
approaches, do not produce energy minima corresponding to localised charge (even though CCSD does)
\cite{Phun24}
while higher-level multi-reference MRCI+Q calculations do.\cite{dmp0}
The existence of a localised charge distribution in the DMP$^+$ cation has become a controversial topic.\cite{wong2018,cheng2018reply,KauppNatComm,Phun24}

Recently, Kaupp and co-workers\cite{KauppNatComm} presented a comparison of state-specific and state-averaged multireference calculations, demonstrating that state-averaged calculations are in agreement with coupled cluster calculations in that an energy minimum corresponding to a localised charge distribution is not produced. They suggest that the state-specific MRCI+Q method used in reference \citenum{dmp0} leads to artificial symmetry-breaking.
For the 3s Rydberg excited state of DMP, however, calculations using the equation-of-motion coupled cluster method produce an energy minimum corresponding to a localised hole separated by an energy barrier from the global minimum where the hole is delocalised.
The binding energy of the Rydberg electron is larger when the hole is more localised and if the energy surface is sufficiently flat in that region, this can induce an energy minimum for a localised hole.  

The large spatial extent of the Rydberg orbital provides a challenge when calculations are carried out using atomic basis sets because of the need to include diffuse enough basis functions. The Rydberg orbital can be artificially confined if sufficiently diffuse basis functions are not included, as has been demonstrated recently in calculations using a real space grid representation.\cite{Sigurdarson23}
The DMP$^+$ cation and the Rydberg excited DMP thus represent significant challenges for electronic structure methods, whether they are based on wave function or density functional approaches.



Here, we report results of density functional calculations of DMP$^+$ and 3s Rydberg DMP, comparing functionals of all rungs of Jacob's ladder: LDA, GGA, meta-GGA, hybrid-GGA, hybrid meta-GGA, and double hybrids. Additionally, 
range-separated hybrid functionals such as $\omega$B97X-V and $\omega$B97M-V 
and the recent machine-learned local hybrid DM21 functional is applied (as well as its other variants DM21m, DM21mc and DM21mu).\cite{dm21}
Finally, the effect of full PZ-SIC applied to PBE is re-evaluated with tighter convergence criteria than were used previously\cite{dmp-nat} and comparison made with results using the downscaled correction, PBE-SIC/2.



In the ground state, the six-membered ring of the DMP molecule, consisting of two nitrogen atoms separated by two carbon atoms on each side,  has the chair form with the two methyl groups attached to the nitrogen atoms in an equatorial position.
Two dihedral angles are used to characterise the structure of the molecule:
angle $d_1$ defined by atoms \{C1,N2,C3,C4\} 
and angle $d_2$, defined by atoms \{C5,N6,C7,C8\}.
The minimum energy structure in the ground electronic state has  
($d_1$,$d_2$)=(175°,-175°).

Figure 1 shows the spin density and illustrates the definition of the dihedral angles for the two types of structures of the cation, DMP$^+$,
as well as energy surfaces obtained with the BHLYP,
DM21 and PBE(.32) functionals.
The BHLYP energy surface is obtained by fixing the dihedral angles and minimising the energy with respect to the remaining degrees of freedom.
In the structure corresponding to the global energy minimum, labeled D, the charge is 
evenly distributed over the two nitrogen atoms, and the methyl groups are in an axial position, giving dihedral angles  
(90°,-90°).
In the two higher energy, local minima, labeled L, the charge is localised on one of the nitrogen atoms and the dihedral angles are 
ca. (130°,-170°) and (170,-130°).
The BHLYP functional 
has a weight of 0.50 on FE,
as does the original hybrid functional developed by Becke,
\cite{BHLYP} 
and
this energy surface has been reported previously in reference \citenum{dmp0}. 
It is reproduced here 
because the obtained atomic coordinates  
are used in the 
calculations shown in figures 1(b) and 1(c).

The neural network generated
DM21 functional is trained to give piece-wise linear energy dependence for both fractional charge and spin.\cite{dm21}
It can, therefore, be expected to give an accurate estimate of the balance between localised and delocalised charge distribution.
Since analytical atomic forces are not available for this functional, the atomic coordinates obtained from the BHLYP calculation shown in figure 1(a) are used to generate the points on the DM21 energy surface shown in figure 1(b).  
Unlike the BHLYP results, local energy minima corresponding to localised charge are not present, but 
energy valleys up from the delocalised state minimum towards those regions of the energy surface are evident.

\begin{figure}[H]
\includegraphics[scale=0.68]{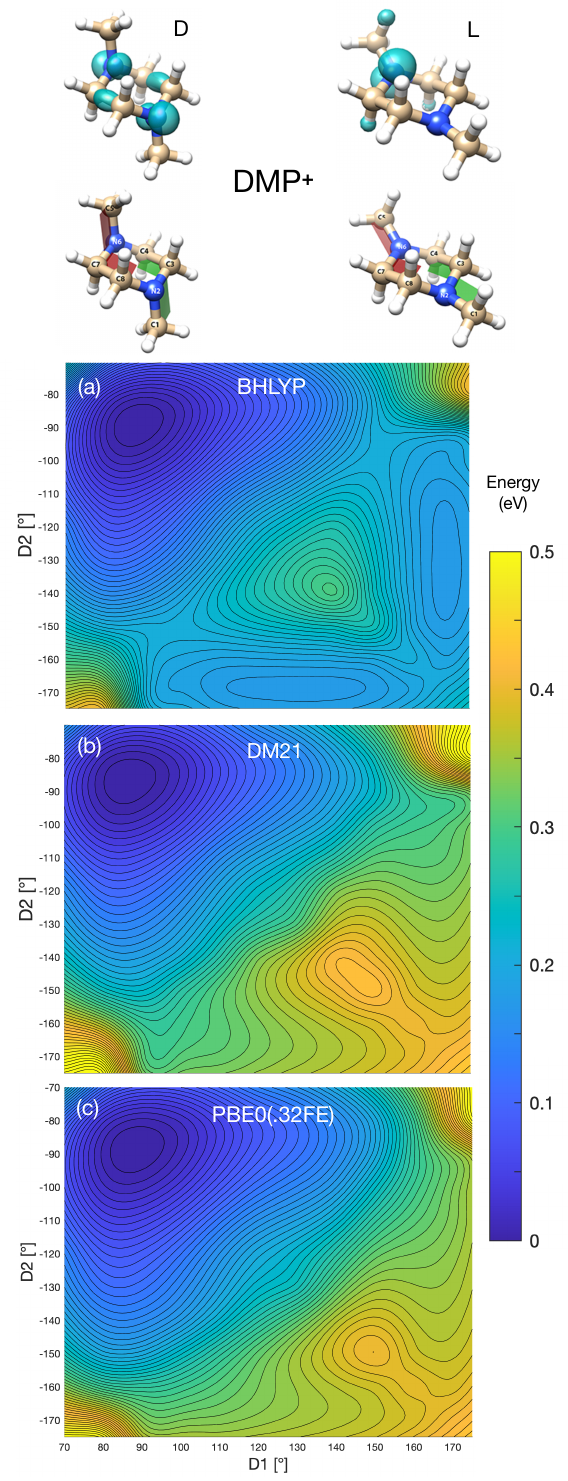}
\caption{
Molecular structure, spin density rendered at 0.01 electron/Å$^3$, and energy surfaces of DMP$^+$ using atomic coordinates obtained from BHLYP calculations.
Top:  Delocalised (left) and localised (right) charge structures, with
planes defining the two dihedral angles of the six-membered ring, $d_1$ 
and $d_2$.
%
(a): Results for the BHLYP functional
where the weight on Fock exchange is 0.50. The atomic structure used for generating the energy surface is obtained by fixing the two dihedral angles for a grid of values and minimising the energy with respect to the remaining degrees of freedom. 
%
The molecular structures shown on top are placed directly above the corresponding energy minima.
(b): Results for the neural network DM21 functional\cite{dm21} of a local hybrid form. Here, only the delocalised minimum is present, but energy valleys stretching towards the regions of the localised charge are evident.
(c): Results for the PBE0 functional\cite{PBE0} where the weight on Fock exchange is increased from the usual value of 0.25 to 0.32 in order to  roughly match the DM21 energy surface shown in (b). 
%
%
}
\label{pes-mrci}
\end{figure}

Figure 1(c) shows the energy surface generated using the global hybrid PBE0 functional\cite{PBE0} where the weight of FE has been increased 
from the usual value of 0.25 to 0.32,
referred to as PBE0(.32FE), in order to roughly match the DM21 energy surface. This shows that a fairly good match can be obtained for the DM21 results
with a global hybrid. The optimal value obtained for the FE weight is similar to what has been used previously for mixed-valence 
systems, as mentioned above.
If the weight on FE is increased further, to 0.50, local energy  minima corresponding to localised charge, analogous to those found for the BHLYP function do form, as discussed below.

%


Figure 2 shows four energy surfaces generated with different functionals and including energy minimisation with respect to the atomic coordinates while the two dihedral angles are kept fixed.
In figure 2(a) the energy surface for the
PBE functional,\cite{pbe_0,pbe_1} is shown. It is of the generalised gradient approximation (GGA) form and 
is the most commonly used functional in calculations of condensed phase systems.
Figure 2(b) shows the energy surface for the PBE0(.32) global hybrid functional, which we take to be our best estimate since it reproduces the DM21 results quite well (see figure 1).
A comparison with the energy surface in figure 1(c) shows the effect of structural relaxation with the PBE0(.32) functional.
The PBE energy surface is very different from the PBE0(.32) energy surface, which we take to be our best estimate. There is a hint of an energy valley along the diagonal instead of the regions corresponding to the localised charge distribution. 
This can be ascribed to the self-interaction error in the PBE functional, as illustrated below.
The PBE(.32) energy surface shows energy valleys in the region of localised charge, but less pronounced than in figure 1(c) where the atomic coordinates are taken from the BHLYP calculation.

Figures 2(c,d) show results for DMP$^+$ when explicit SIC as proposed by Perdew and Zunger\cite{sic} is applied to the PBE functional.
This is an orbital-by-orbital correction and therefore is most applicable for an independent electron system.
It tends to overcorrect, for example in calculations of atomisation energy of molecules, but a scaling of the correction by 0.50 has been shown to give accurate results. 
Figure 2(c) shows the energy surface when SIC scaled by 0.50 is applied to the PBE functional.
A qualitatively similar energy surface is obtained as for PBE(.32). Again, there is only one energy minimum, the one corresponding to the delocalised charge.  However, when full SIC is applied, as shown in figure 2(d), clear minima are present corresponding to localised charge 
in regions where analogous minima are present in the BHLYP surface (shown in figure 1(a)). 
A similar energy surface is obtained when a weight of 0.50 is used in the PBE0 functional, referred to as PBE0(.50). The hybrid functionals with FE weight of 0.50 and the PBE functional with full SIC applied give similar energy surfaces. The energy barrier  between the minima corresponding to localised and delocalised charge are also similar, as shown in figure 3 and listed in table 1.

\begin{figure}[H]
\includegraphics[scale=0.37]{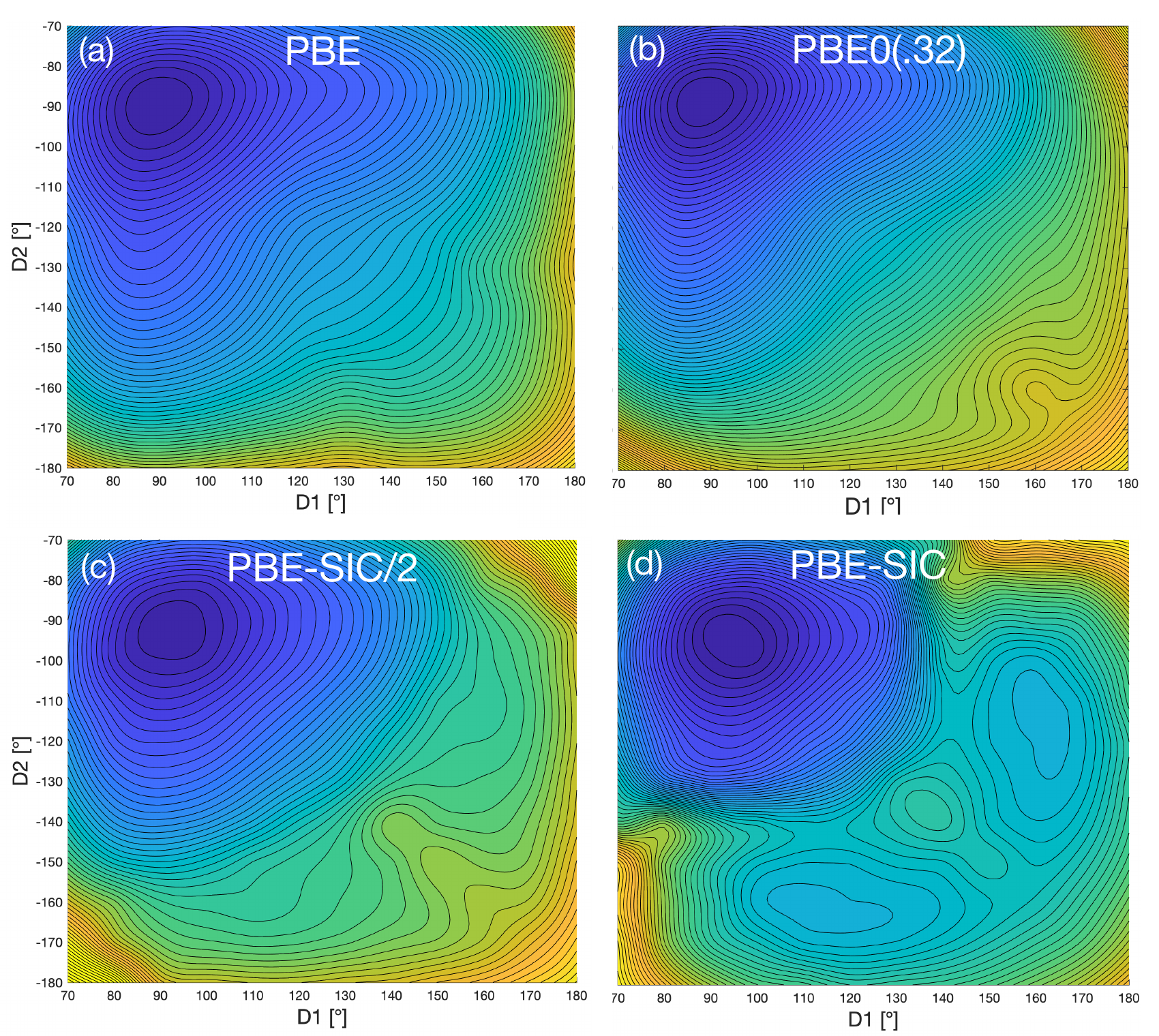}
\caption{
Energy surfaces for DMP$^+$ where the energy is minimised with respect to atomic coordinates for each functional while keeping the two dihedral angles fixed.   
(a) PBE generalised gradient approximation functional, 
(b) PBE0(.32) global hybrid functional with FE weight of 0.32,
(c) PBE-SIC/2 functional where self-interaction correction\cite{sic} scaled by 0.50 is applied to PBE,
(d) PBE-SIC functional where full self-interaction correction is applied to PBE.
PBE-SIC/2 gives results similar to the PBE0(.32) functional in that minima corresponding to localised charge are not present but 
energy valleys stretching into the localised charge region are evident. PBE-SIC gives results similar to hybrid functionals with 0.50 weight on Fock exchange, BHLYP and PBE(.50),
see also minimum energy paths in figure 3.
The colour scale is the same as in figure 1.
}
\label{pes-cation}
\end{figure}
\begin{figure}[H]
\includegraphics[scale=0.32]{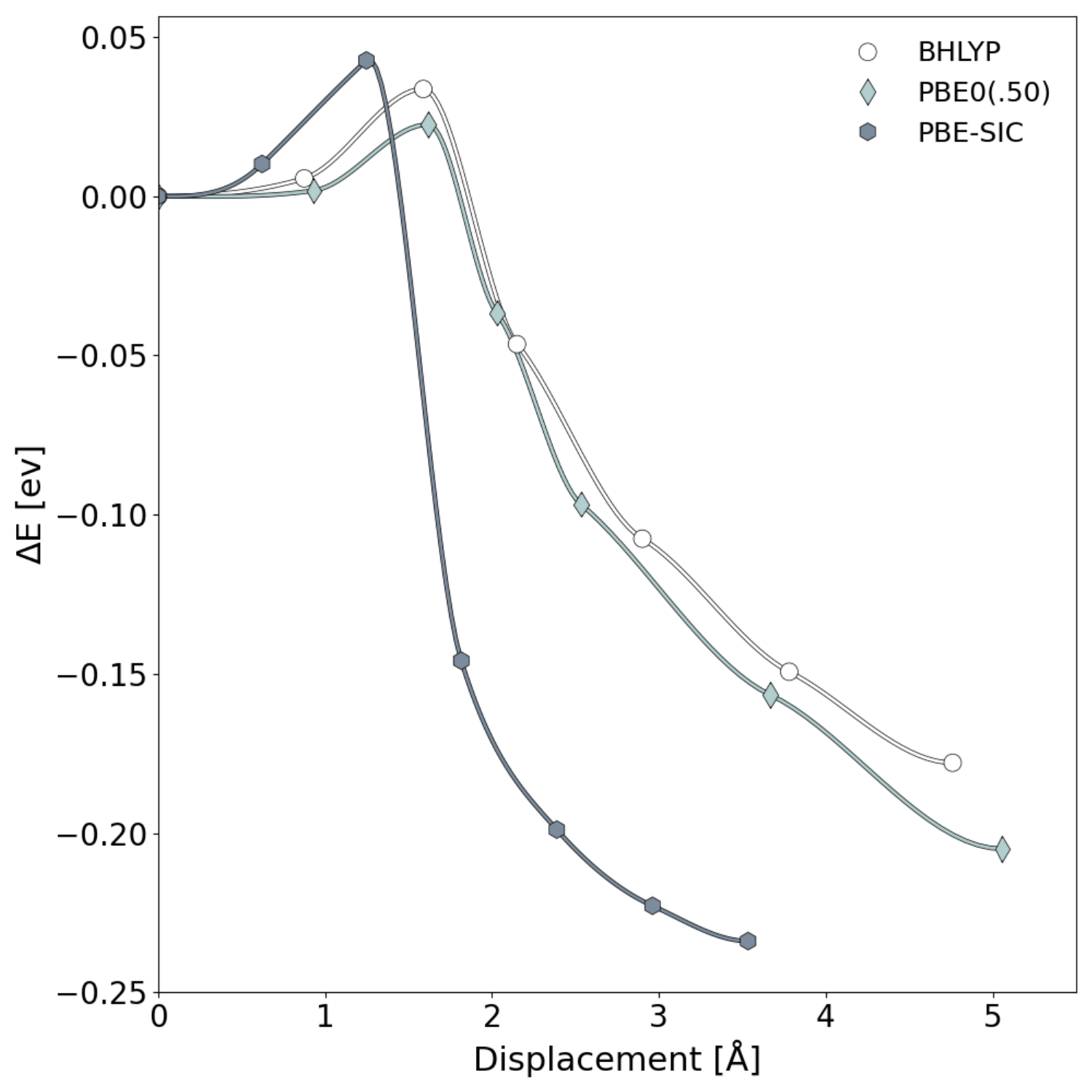}
\caption{
Energy along the minimum energy path between the energy minima corresponding to localised and delocalised charge of DMP$^+$ 
calculated using the BHLYP, PBE0(.50) and PBE-SIC functionals.
The total path length differs mainly because of a rotation of the methyl groups which, however, correspond to only small change in the energy.
Similar energy barrier is obtained with some of the double hybrid functionals, see table 1.
}
\label{neb-xc}
\end{figure}
\begin{table}[H]
  \caption{
Energy barrier, E$_{b}$, for the transition from the local energy minimum corresponding to localised charge (when present) to the global minimum corresponding to charge delocalised over both nitrogen atoms, and the
corresponding change in energy, $\Delta$E, of the DMP$^+$ ion. 
The values for the functional calculations are obtained from nudged elastic band calculations without any constraints on the atomic coordinates as in the minimum energy paths shown in figure 3 (not from the energy surfaces where the two dihedral angles are fixed).
The results from the wave function based multi-reference configuration interaction, MRCI+Q, calculations are taken from reference \citenum{dmp0}.
  }
  \label{tbl:stationary}
  \begin{tabular}{lccc}
    \hline
    Functional   & E$_{b}$ [eV] & $\Delta$E [eV]  \\
    \hline
    PBE0(.50)
                               & 0.022          & -0.21       \\
    BHLYP               & 0.033          & -0.18     \\
    PBE-SIC             & 0.046          & -0.23      \\
    B2GPPLYP         & 0.002          & -0.48      \\
    DSD-BLYP          &0.013           & -0.56   \\
    DSD-PBEP86    & 0.017          & -0.49   \\
    MRCI+Q\cite{dmp0}            & 0.054          & -0.38      \\
    \hline
  \end{tabular}
\end{table}

Several other functionals are tested. The full list is given in table 2 where it is specified whether or not a local minimum corresponding to  localised charge is found.  
In each case, a minimisation of the energy with respect to atom coordinates is carried out using a careful optimisation based on 
accelerated steepest descent starting from an initial configuration in the region corresponding to localised charge. 
Analysis of the spin density is used to determine whether localised or delocalised charge is obtained in the end. 
For all the GGA and meta-GGA functionals, such calculations only produce the delocalised charge energy minimum. This includes the SCAN and r$^2$SCAN meta-GGA functionals. Global hybrid functionals, such as B3LYP and PBE0 
(with the usual FE weight of 0.20 and 0.25, resp.), 
also only produce an energy minimum corresponding to delocalised charge. 
Same goes for the range-separated hybrid functionals.

Calculations are carried out for four double-hybrid density functionals that incorporate MP2 correlation as well as FE. 
They typically include higher weight on FE than regular hybrids.
The B2GPPLYP (with 0.65 FE),\cite{DSD-PBEP86} B2PLYP (with 0.69 FE)\cite{DSD-PBEP86} and DSD-BLYP (with 0.70 FE)\cite{DSD-BLYP} functionals give local energy minima corresponding to localised charge as well as the global minimum for the delocalised charge, while the B2PLYP (with 0.53 FE)\cite{B2PLYP} functional only gives the delocalised charge energy minimum, see table 2. 
The presence of localised charge energy minima for some of the double hybrids is clearly associated with the large weight on FE which is not compensated sufficiently by the addition of MP2 correlation.
The energy of the localised charge minimum, when present, is quite high as shown in table 1.
The energy barrier for a transition from the localised to the delocalised charge minimum of the DSD-PBEP86 and DSD-BLYP functionals, 
0.017 eV and 0.013 eV, is close to that obtained with the PBE0(.50) functional. The B2GPPLYP functional produces a smaller barrier, see table 1.   
The more recently developed local hybrid functionals\cite{kaupp2019_LHs} are interesting alternatives but have not been applied here.

Figure 4 shows energy surfaces obtained for the 3s Rydberg state of DMP using the variational excited state calculations.  The atomic coordinates are optimised by energy minimisation keeping the two dihedral angles fixed.  The energy surface for the PBE functional,
shown in figure 4(a), does not show signs of local energy minima corresponding to a localised hole, 
only a minimum where the hole is distributed equally over the two nitrogen atoms. 
There is some difference from the energy surface of the cation shown in figure 2(a),
mainly increased energy along the diagonal where the two dihedral angles are equal.

\begin{table}[H]
  \caption{
List of the various density functionals tested and an indication whether a local energy minimum corresponding to 
a localised charge, L, is found on the energy surface of the DMP$^+$ ion.
B3LYP(.50) and PBE0(.50) denote functionals where the weight of Fock exchange has been increased to 0.50 from the 
usual value of 0.20 and 0.25, respectively.
PBE-SIC/2 denotes self-interaction correction\cite{sic} scaled by 0.50 and applied to the PBE functional, while
PBE-SIC denotes full correction.
  }
  \label{tbl:multi}
  
  \begin{tabular}{lcccc}
    \hline
Functional     & L minimum &    \\
 \hline
    {\bf{GGA:}} \\
    \ \ \ BLYP\cite{BLYP_Becke1988,BLYP_Lee1988} & -             \\
    \ \ \ BP86\cite{BP86_Becke1988,BP86_Perdew1986} & -              \\
    \ \ \ OLYP\cite{OLYP_Handy2001,BLYP_Lee1988} & -             \\
    \ \ \ PBE\cite{pbe_0,pbe0_1}  & -              \\
    \ \ \ PW91\cite{PW91_Perdew1992} & -              \\
    \ \ \ RPBE\cite{RPBE_Hammer1999} & -             \\
\hline
{\bf{meta-GGA:}} \\
     \ \ \ B97M-D3BJ\cite{B97M-D3BJ_Lin2013} & -        \\
     \ \ \ TPSS\cite{TPSS_Tao2003}      & -             \\
     \ \ \ revTPSS\cite{revTPSS_Perdew2009}   & -          \\
     \ \ \ r2SCAN\cite{r2SCAN_Furness2020}      & -        \\
\hline
{\bf{Global hybrids:}} \\
     \ \ \ TPSS0\cite{TPSS0_Staroverov2003}     & -            \\
     \ \ \ TPSSh\cite{TPSS0_Staroverov2003}            & -             \\
     \ \ \ B3LYP\cite{B3LYP_Stephens1994}            & -              \\
     \ \ \ PBE0\cite{PBE0}             & -                \\
     \ \ \ PW6B95\cite{PW6B95_Zhao2005}           & -                \\
     \ \ \ BHLYP          & \checkmark            \\
     \ \ \ B3LYP(.50) & \checkmark      \\
     \ \ \ PBE0(.50)  & \checkmark        \\
\hline
{\bf{Self-interaction correction:}} \\
      \ \ \ PBE-SIC/2\cite{Klupfel12}      & -            \\
    \ \ \ PBE-SIC\cite{sic}      & \checkmark             \\
\hline
{\bf{Range-Separated hybrids:}} \\
     \ \ \ CAM-B3LYP\cite{CAM-B3LYP_Yanai2004}         & -               \\
     \ \ \ $\omega$B97M-D3BJ\cite{wB97M-D3BJ_Lin2016} & -          \\
     \ \ \ $\omega$B97X-D3BJ\cite{wB97M-D3BJ_Lin2016} & -               \\
\hline
{\bf{Double hybrids:}} \\
      \ \ \ B2PLYP\cite{B2PLYP}          & -                  \\
      \ \ \ B2GPPLYP\cite{B2GPPLYP}   & \checkmark        \\
      \ \ \ DSD-BLYP\cite{DSD-BLYP}   & \checkmark         \\
      \ \ \ DSD-PBEP86\cite{DSD-PBEP86} & \checkmark          \\
\hline
{\bf{Neural network:}} \\
    \ \ \ DM21\cite{DM21_Kirkpatrick2021}          & -                \\
    \ \ \ DM21m\cite{DM21_Kirkpatrick2021}      & -                \\
    \ \ \ DM21mc\cite{DM21_Kirkpatrick2021}     & -                \\
    \ \ \ DM21mu\cite{DM21_Kirkpatrick2021}     & -                \\
 \hline
 \end{tabular}  
\end{table}



\begin{figure}[H]
\includegraphics[scale=1.1]{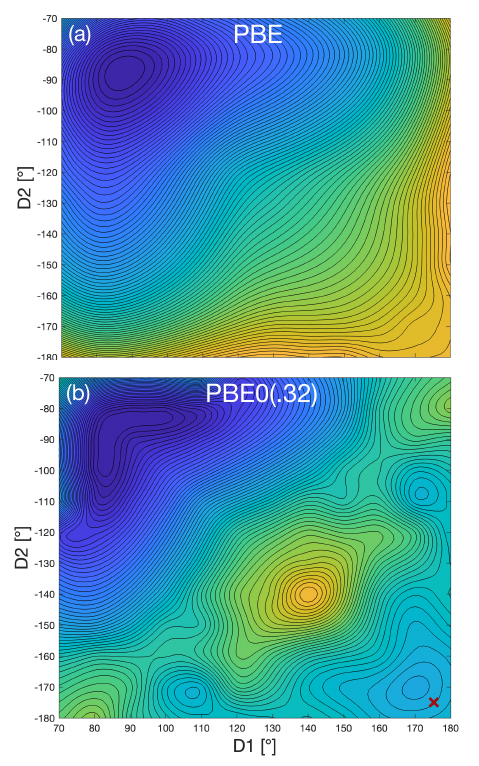}
\caption{
Energy surfaces for the 3s Rydberg state of DMP.
(a) Calculations using a generalised gradient approximation functional, PBE.
Only the minimum corresponding to the delocalised charge is present and there is no sign of local minima corresponding to localised charge.
(b) Calculations using the PBE0 functional with a weight of 0.32 for Fock exchange. Two different types of minima are now present corresponding to localised hole, in addition to the global minimum corresponding to a delocalised hole.
The ground state minimum energy structure is indicated by a red +.
The colour scale is the same as in figure 1.
}
\label{pes-Rydberg}
\end{figure}

The calculation using the PBE0(.32) hybrid functional, however, shows dramatic difference between the energy surface of the Rydberg state, shown in figure 4(b), and that of the cation, shown in figure 2(b). Two types of energy minima appear corresponding to a localised hole.
The deeper one, located on the diagonal, corresponds to molecular structure analogous to the optimal ground state structure, where the two methyl groups are in an equatorial position.  
The dihedral angles have changed slightly from the ground state values by ca. 5°.
Even though the two dihedral angles are equal in this structure, it is not symmetric as the two nitrogen atoms are inequivalent. 
The shallower energy minimum where the dihedral angles have different values is qualitatively similar to the localised charge structure, L, of the cation. The energy barrier between these two minima represents the activation energy for the equatorial to axial transition of the methyl group where the hole resides. 
%
The presence of two types of localised hole structures shown by our calculations is consistent with the interpretation of the experimental measurements in terms of a multitude of conformeric structures with a localized hole.\cite{Sanghamitra13} 
The surprisingly large difference between the energy surface of the Rydberg state and that of the cation is consistent with the calculations 
reported by Kaupp and coworkers where localisation of the hole in the Rydberg state was found while the cation charge did not localise.\cite{KauppNatComm}
Further analysis of the 3p and 3s Rydberg states will be presented at a later time. The main emphasis here is on the cation.



The balance between localised and delocalised charge distribution of the DMP$^+$ ion and 3s Rydberg excited state of DMP represents a sensitive and challenging test of electronic density functionals. 
This is a rare case where both a delocalised and localised charge or hole can be present on the same adiabatic energy surface.
While it is clear from experimental measurements that a localised hole exists for the Rydberg DMP,\cite{dmp-nat} 
a localised charge state most likely does not exist for the DMP$^+$ ion.  The energy surface of DMP$^+$ must, however, be relatively flat in the region of the localised charge in order for the Rydberg electron to induce such a minimum since the Rydberg electron is highly dispersed in space. 
The various density functionals tested here give quite different shape for the energy surface of DMP$^+$.
Three of the four double hybrid functionals tested here, predict the presence of a local, symmetry broken minimal energy structure 
for the DMP$^+$ ion corresponding to a localised charge.
Lower on the Jacob's ladder, the GGA functionals, such as PBE, do not produce a localised state and the PBE energy surface shown in figure 2(a) does not even display energy valleys in the region of the localised charge. On the contrary, a hint of a valley lies along the diagonal. The shape of the surface changes by adding Fock exchange in a hybrid functional and the surface produced with the PBE0(.32) functional where the weight on FE is 0.32 shows a hint of valleys in the direction of the symmetry broken localised charge structures. 
The functional gives a fairly good match with the machine learned DM21 functional where the 
input data base includes data on fractional charge and spin.
In order to produce additional energy minima on the DMP$^+$ energy surface, the weight on FE needs to be increased further. 
With a weight of 0.50 as in either the PBE0(.50) or the BHLYP functionals, distinct energy minima are formed corresponding to localised charge. 
This evolution of the shape of the energy surface as a function of increased weight on FE is a reflection of the extent to which the self-interaction in the estimate of the classical Coulomb interaction based on the total electron density is canceled out and even eventually overcorrected. 
The FE term in the hybrid functionals contains self-interaction of opposite sign to the one in the classical Coulomb interaction, as well as the infinite range exchange terms.
The self-interaction in the classical Coulomb interaction favours delocalisation while FE favours localisation.
In the extreme case of pure HF, only the localised state is present for the DMP$^+$ ion.\cite{dmp-nat}  

The role played by the self-interaction error can be seen more clearly by making an explicit self-interaction correction to the PBE functional. The orbital-by-orbital correction proposed by Perdew and Zunger\cite{sic} is exact for a system with only one electron, but does not take many-body effects into account. 
The full correction is needed in order to obtain the correct -1/$r$ long range form of the effective potential of an electron, an important feature for accurate representation of diffuse orbitals in Rydberg excited states and loosely bound anions.\cite{Gudmundsdottir13,Zhang16}
But, it has been shown,\cite{Jonsson11,Klupfel12} and justified by reference to the adiabatic connection formula, as in the original hybrid functional of Becke,\cite{BHLYP} that the correction term should be scaled by 0.50. This gives significantly improved results for atomisation energy of molecules and band gap of solids. It has also been shown to give good results for the delicate balance between different electronic energy surfaces of the Mn dimer.\cite{Ivanov21} 
One of the interesting results from the present analysis of the energy surface of the DMP$^+$ ion is that full correction applied to PBE gives results that are similar to having a weight of 0.50 on FE in the PBE0(.50) and BHLYP functionals, while scaling the correction by 0.50 is closer to the machine learned DM21 functional in terms of the qualitative shape of the energy surface.     

The  energy barrier for a transition from the localised state to the delocalised state of DMP$^+$ turns out to be in the range of 0.02 to 0.045 eV for the functionals where a localised state minimum is present, namely the DSD-PBEP86 double hybrid functional, the BHLYP and PBE0(.50) global hybrid functionals, and the PBE functional with full self-interaction correction applied, PBE-SIC,
as shown in table 1. 
This value presented of the activation energy for PBE-SIC is lower than a previously reported estimate\cite{dmp-nat} mainly because the calculation is here carried out to a tighter convergence. The convergence tolerance for the force acting on the climbing image in the CI-NEB calculations is 0.01 eV/Å in the present case, while it was 0.05 eV/Å in the previous calculation (a default setting in the GPAW software).
Improvements in the implementation of PZ-SIC in GPAW in recent years\cite{Lehtola16,Ivanov21a,Ivanov21b,gpaw2} has made it possible to reach better convergence. 
There, the full PZ-SIC was applied in order to obtain the correct -1/$r$ long range form of the effective potential as the goal was in part to obtain an accurate estimate of the Rydberg electron binding energy.\cite{dmp-nat} 
The scaling down of the PZ-SIC by a factor of 0.50, i.e. the PBE-SIC/2 functional, eliminates the local energy minima corresponding to localised charge of DMP$^+$. 
Optimally, the Rydberg orbital should be subject to full PZ-SIC while a scaled-down correction should be applied to the rest of the orbitals.
The difference between the two methods for addressing the self-interaction error in semi-local Kohn-Sham functionals, the inclusion of FE in hybrid functionals and the explicit PZ-SIC, is mainly the off-diagonal terms in the infinite range FE. They ensure the hybrid functionals are unitary invariant and thereby simplify the computational effort while the application of PZ-SIC makes the functional explicitly dependent on the orbital densities rather than just the total electron density. 

Whenever possible, tests of density functionals are carried out against results of high-level wave function based approaches. The DMP molecule is, however, large enough to make this a significant challenge.
For reference, table 1 also lists results obtained with wave function based multi-reference configuration interaction, MRCI+Q, reported in 
reference \citenum{dmp0}. As explained there, an energy surface was generated using atomic coordinates obtained from BHLYP calculations because the MRCI+Q calculations require too large computational effort for the structural relaxation. The value of the energy can, therefore, be expected to be a bit higher in general. Clear minima corresponding to the localised state are present on the energy surface presented in reference \citenum{dmp0} and the overall shape somewhat similar to that obtained with PBE0(.50) and PBE-SIC. 
From MRCI+Q calculations using the fully relaxed atom coordinates and first order saddle point obtained by BHLYP, the energy of the localised state is found to be 0.38 eV with respect to that of the delocalised state and the energy barrier 0.054 eV, a bit higher than for the density functionals that include FE with a weight of 0.50. 
Subsequent state averaged calculations carried out by Kaupp and coworkers,\cite{KauppNatComm} however, do not show an energy minimum corresponding to a localised charge and it is argued there that the 
state-specific CASSCF calculation
which underlies the MRCI method induces a cusp that leads to an artificial energy barrier in the 
MRCI+Q calculations. 

The energy surface for the 3s Rydberg excited state of DMP obtained with the PBE0(.32) functional turns out to be significantly different from the energy surface of the DMP$^+$ ion. 
Minima corresponding to localised charge are present on the Rydberg surface, while they are absent on the cation surface.
This is an interesting result since the energy surface of a Rydberg state is often assumed to be quite similar to that of the corresponding cation.
\cite{Stankus19,Yong21}
This result is in agreement with the calculations of Kaupp and coworkers.\cite{KauppNatComm}
For the PBE functional the difference between the Rydberg and cation surfaces is not as large as for the hybrid functional and neither one shows minima corresponding to localised charge or hole in our calculations.
This means that the self-interaction error in the PBE functional needs to be removed to a sufficient extent in order for the presence of the Rydberg electron to induce an energy minimum corresponding to a localised hole. 
Kaupp and coworkers, however, did report a very shallow energy minimum for the PBE functional in time-dependent DFT calculations.\cite{KauppNatComm}

The PBE0(.32) energy surface for the Rydberg excited state of DMP appears to be qualitatively consistent with the experimental measurements presented in reference \citenum{dmp0} in that energy minima for a localised hole are present, but here two types of such minima are found.
This and other aspects of the Rydberg state calculations 
as well as more detailed comparison with the experimental data
will be the subject of future studies.



In conclusion, the energy surfaces of the DMP$^+$ ion and the 3s Rydberg excited state of DMP show well the balance between delocalised and localised charge distribution and
provide valuable tests for the 
weight of FE in hybrid functionals, or alternatively 
the extent to which explicit orbital-based self-interaction correction should be applied. 
The inclusion of FE  
in the PBE0 functional with a sufficiently large weight, such as 0.32, stabilises localisation enough for the Rydberg energy surface to produce minima corresponding to a localised hole. 
The DMP$^+$ energy surface for PBE0(.32), however, does not show evidence of localised charge. This qualitative difference between the cation surface and the Rydberg surface is noteworthy because it is generally assumed that the energy surface of a Rydberg excited state is similar to that of the cation. 
When the weight of FE is increased to 0.50, as in BHLYP and PBE0(.50), also the DMP$^+$ energy surface supports a localised charge.
Similar DMP$^+$ energy surface is obtained when the full PZ-SIC is applied.  However, when the SIC is scaled down by a factor of 0.50, which can be justified from the adiabatic connection formula as in the original hybrid functional of Becke, 
the energy minima corresponding to localised charge disappear. A rough correspondence emerges between a weight of 0.50 on FE and full SIC on the one hand, and a weight of {\it ca.}\,0.32 on FE in PBE0 and a downscaling of SIC by 0.50 on the other hand. 
It is clear that too large weight on FE or SIC leads to artificial symmetry breaking.
Given the large data base for training and the emphasis on accurate representation of the variation of the energy as a function of fractional charge, it is likely that the machine learned functional DM21 gives 
a good
estimate of the 
balance between localisation and delocalisation.
While it does not produce minima on the energy surface of DMP$^+$ corresponding to localised charge, it does show clear energy valleys towards those regions.
Similar shape of the energy surface can be obtained by increasing the weight on FE to a value of 0.32 in the PBE0 functional.
This is reassuring since a value in this range has been found to be optimal in comparison with results of high-level quantum chemistry calculations of various systems in order to obtain the right balance between localisation and delocalisation.
%
The range-separated hybrid functionals used here do not produce a metastable localised state for DMP$^+$, but most of the tested double-hybrid functionals do.


\section{Methods}

Most of the DFT calculations are carried out using the aug-cc-pVDZ\cite{augccbasis0,augccbasis1} basis set as implemented in the ORCA quantum chemistry software.
\cite{orca5} 
However, the self-interaction corrected PBE calculations (both full correction, PBE-SIC, and scaled correction, PBE-SIC/2), 
as well as some tests of the Rydberg excited state calculations using PBE-SIC/2,
are carried out using the GPAW software and a grid based representation of the valence electron orbitals combined with the projector augmented wave representation of the effect of the inner electrons.\cite{gpaw2}
In order to test the consistency of the two approaches, 
and the convergence with respect to the atomic basis set,
some of the calculations
are done in both ways and only insignificant difference is found between the resulting energy surfaces.
Calculations using the DM21 functionals\cite{dm21} are carried out with the PySCF software\cite{pyscf0,pyscf1,pyscf2}
and the aug-cc-pVDZ basis set.

The Rydberg excited state calculations are carried out using a variational
orbital optimisation
method where an excited state is obtained by converging on the corresponding saddle point on the electronic energy surface\cite{Levi2020a,Levi2020b,Schmerwitz23}
as implemented in ORCA version 6.0.0. 
The calculations are carried out for the triplet as well as mixed singlet and spin purification\cite{Ziegler1977} 
$E^{s} = 2E(\uparrow\downarrow) - E(\uparrow\uparrow),$ is carried to get an improved estimate of the pure singlet energy surface.
This approach has been proven to work well in calculations of Rydberg excited states of molecules, including open shell singlets,\cite{Ivanov21b,Schmerwitz22,Sigurdarson23} as well as for excited defect states in solids.\cite{Ivanov23} 

Calculations of the energy barrier between the localised state, when present, and the delocalised state are carried out using the
climbing image nudged elastic band (NEB) method\cite{cineb,nebtang,neborca} including 5 intermediate images. 
The calculations are considered to be converged when the magnitude of the force on each atom in the climbing image drops below  0.01 eV/Å.

Energy surfaces characterising the cation are generated by producing a grid of 
values for the two dihedral angles, $d_1$ and $d_2$, of the six-membered ring (illustrated in figure 1) and minimising the energy with respect to the remaining degrees of freedom in the position of atoms, except in the DM21 calculations 
where the atomic coordinates obtained with BHLYP are used 
because analytical atomic forces are not available for the DM21 functional.
Also, the same coordinates are used in the PBE0(.32) calculations in figure 1 (but not in figure 2) since the goal there is to compare with the DM21 results.
In the PBE0(.32) functional, the weight of Fock exchange is increased to 0.32 from the more commonly used value of 0.25.\cite{pbe_0,pbe_1}
The calculated points are
interpolated with a biharmonic spline interpolation using Matlab \cite{matlab} to produce smooth energy surfaces. 
For all the functionals used here, the global minimum on the energy surface of DMP$^+$ corresponds to a delocalised charge. 
Localised charge is also found for some of the functionals, i.e. higher energy local minima on the energy surface where the 
positive charge and the spin density is centred on one of the two N-atoms, 
as illustrated in the ToC graphic.

%
\section{Data and Software Availability}
Data related to the results presented in this article and instructions on the generation thereof are available upon request from the corresponding author.
%


\begin{acknowledgement}

This work was funded by the Icelandic Research Fund, grant no. 2511544 and the University of Iceland Research Fund.
M.G. acknowledges financial support from the SONATA research grant from the National Science Centre, Poland (Grant No. 2023/51/D/ST4/02796).
The calculations were performed on resources provided by the Icelandic Research e-Infrastructure facility.
We thank Alec E. Sigurdarson, Vilhjálmur Ásgeirsson, Gianluca Levi, Aleksei Ivanov and Peter Weber for helpful discussions.

\end{acknowledgement}

\bibliography{revisedmanusctipt}

\end{document}